\documentclass[sigconf, nonacm]{acmart}
\AtBeginDocument{%
  }

\copyrightyear{2026}
\acmYear{2026}
\setcopyright{cc}
\setcctype{by}
\acmConference[FORGE '26]{2026 IEEE/ACM Third International Conference on AI Foundation Models and Software Engineering}{April 12--13, 2026}{Rio de Janeiro, Brazil}
\acmBooktitle{2026 IEEE/ACM Third International Conference on AI Foundation Models and Software Engineering (FORGE '26), April 12--13, 2026, Rio de Janeiro, Brazil}
\acmPrice{}
\acmDOI{10.1145/3793655.3793728}
\acmISBN{979-8-4007-2477-0/2026/04}

\usepackage{subcaption}
\usepackage{multirow}
\usepackage{enumitem}
\usepackage{soul}

\begin{document}

\title{Impacts of Generative AI on Agile Teams' Productivity: A Multi-Case Longitudinal Study}



\author{Rafael Tomaz}
\email{rafaelstomaz@gmail.com}
\orcid{0009-0009-1435-4372}
\authornotemark[1]
\affiliation{%
  \institution{Pontifical Catholic University of Rio de Janeiro}
  \city{Rio de Janeiro}
  \country{Brasil}
}

\author{Paloma Guenes}
\email{pguenes@inf.puc-rio.br}
\orcid{0009-0004-8080-1760}
\affiliation{%
  \institution{Pontifical Catholic University of Rio de Janeiro}
  \city{Rio de Janeiro}
  \country{Brasil}
}
\affiliation{%
  \institution{University of Bari}
  \city{Bari}
  \country{Italy}
}

\author{Allysson Allex Araújo}
\email{allysson.araujo@ufca.edu.br}
\orcid{0000-0003-2108-2335}
\affiliation{%
  \institution{Federal University of Cariri}
  \city{Juazeiro do Norte}
  \country{Brasil}
}

\author{Maria Teresa Baldassarre}
\email{mariateresa.baldassarre@uniba.it}
\orcid{0000-0001-8589-2850}
\affiliation{%
  \institution{University of Bari}
  \city{Bari}
  \country{Italy}
}

\author{Marcos Kalinowski}
\email{kalinowski@inf.puc-rio.br}
\orcid{0000-0003-1445-3425}
\affiliation{%
  \institution{Pontifical Catholic University of Rio de Janeiro}
  \city{Rio de Janeiro}
  \country{Brasil}
}

\renewcommand{\shortauthors}{Tomaz et al.}

\begin{abstract}

\textbf{Context:} Generative Artificial Intelligence (GenAI) tools, such as GitHub Copilot and GPT tools, represent a paradigm shift in software engineering. While their impact is clear, most studies are short-term, focused on individual experiments. The sustained, team-level effects on productivity within industrial agile environments remain largely uncharacterized.   
\textbf{Goal:} This study aims to provide a longitudinal evaluation of GenAI's impact on agile software teams. We characterize its effect on developers' productivity by applying the multi-dimensional SPACE framework.  
\textbf{Method:} We conducted a multi-case longitudinal study involving 3 agile teams at a large technology consulting firm for around 13 months. We collected and compared quantitative telemetry (Jira, SonarQube, Git) and qualitative survey data from historical (pre-adoption) and research (post-adoption) sprints.   
\textbf{Conclusion:} GenAI tools can significantly improve team performance and well-being. Our key finding is a sharp increase in $P$erformance and perceived $E$fficiency concurrent with flat developer $A$ctivity. This suggests GenAI increases the value density of development work, not its volume. This finding validates the necessity of multi-dimensional frameworks like SPACE to capture the true, nuanced impact of GenAI in-situ, which would be invisible to studies measuring $A$ctivity alone.   
\end{abstract}




\maketitle

\section{Introduction}

Foundation Models (FMs) are large AI models trained on vast datasets that can be adapted to perform a wide range of tasks, from language processing to image generation \cite{wiggins2022opportunities}. The emergence of FMs has triggered a structural shift in Software Engineering (SE). Tools, for example, internal GPT tools (an internal, GPT-powered chatbot designed to query documentation and assist in drafting user stories, technical documentation, and Pull Requests) and GitHub Copilot, operationalize these models to assist developers in code generation, refactoring, documentation, and test creation \cite{ebert2023generative, sauvola2024future}. Beyond automation, FMs act as cognitive collaborators, augmenting human reasoning and transforming how teams design, communicate, and deliver software \cite{peng2023impact}. Consequently, their rapid diffusion has also sparked critical questions about how productivity evolves when intelligent assistance becomes embedded in everyday workflows.

At the same time, agile methodologies remain the prevailing paradigm for managing uncertainty and promoting continuous value delivery in SE \cite{gandhira2025adoption}. Agile teams emphasize collaboration, adaptability, and rapid feedback, values that align conceptually with the generative and assistive nature of FM \cite{assalaarachchi2025generative}. Yet, the empirical understanding of how FM-powered tools interact with agile practices remains limited. Recent empirical studies have begun to quantify short-term productivity gains and capture developer perceptions of Generative AI (GenAI) in programming tasks \cite{peng2023impact,murali2024ai,zhang2023practices}. What remains unclear is how the sustained use of FM reshapes the multidimensional experience of software teams over time, particularly in terms of productivity, as teams adapt their workflows and coordination practices. Evidence of such longitudinal and team-level dynamics remains scarce, especially in large IT consulting firms where multiple agile squads operate under intense delivery constraints and share organizational processes. 

To address this gap, we conducted a multi-case longitudinal study with three agile teams from an IT consulting firm recognized as one of the world’s largest professional service networks. The firm operates in more than 150 countries, has more than 300k employees, and delivers an integrated wide range of services to customers across multiple sectors. These environments provide a valuable "laboratory" for investigating the integration of FMs in real software projects, striking a balance between contextual diversity and methodological consistency. Over a one-year period, we compared historical sprints (pre-adoption) with research-period sprints (post-adoption) to evaluate changes in productivity, well-being, and collaboration dynamics following the introduction of an internal GPT tool and GitHub Copilot. The firm provided complete organizational support to enable the researchers to investigate the impact of GenAI. This study integrates quantitative metrics extracted from Jira, SonarQube, and Git repositories with qualitative data from surveys and semi-structured interviews, allowing both within-team and cross-case comparisons that reveal how FM adoption evolves and stabilizes over time.

Our investigation was primarily guided by the SPACE framework \cite{forsgren2021space} because it provides a holistic view of developer productivity that extends beyond output-based metrics. In summary, SPACE defines five dimensions: Satisfaction and well-being ($S$), Performance ($P$), Activity ($A$), Communication and collaboration ($C$), and Efficiency and flow ($E$). To operationalize these dimensions in a structured manner, we applied the Goal–Question–Metric (GQM) paradigm \cite{basili2002tame}, linking the study’s objectives to measurable indicators and survey constructs. This integration provides a structured and replicable approach to studying human–AI collaboration, bridging quantitative analytics with subjective experience data. Furthermore, the research protocol incorporated standardized workshops to establish a common baseline knowledge for all participants regarding prompt engineering and tool usage.

Building on this research design, the study investigates how the use of the internal GPT tool and GitHub Copilot affects the productivity of agile teams in real organizational settings. To the best of our knowledge, this research presents the first in-depth multi-case and longitudinal analysis of FM integration within major Technology Consulting firms. It provides empirical evidence regarding the ongoing impact of Generative AI (GenAI) tools in agile settings, enhancing current research that typically emphasizes short-term or individual application. 

Our main finding indicates that overall team performance and perceived efficiency rose significantly, even though the level of developer activity (\textit{e.g.}, commits and lines of code) remained steady. This suggests that generative AI boosts productivity not by increasing work throughput, but by enhancing the value of each contribution, allowing developers to focus more on complex reasoning and validation instead of routine code writing.


The remainder of this paper is organized as follows. Section 2 reviews the background and related work on foundation models in software engineering. Section 3 details the study design and data collection procedures. Section 4 presents the results, followed by discussion (Section 5) and threats to validity (Section 6). Section 7 concludes the paper.

\section{Background and Related Work}

Generative Artificial Intelligence (GenAI), powered by Foundation Models (FMs), is rapidly influencing software engineering (SE) practices. Tools that integrate Large Language Models (LLMs), such as GitHub Copilot, have moved from novelties to embedded components in developers' daily workflows \citep{sauvola2024future}. The primary promise of these tools is a significant enhancement in team efficiency and output. Initial research has begun to quantify these benefits, sparking a necessary investigation into how these tools fundamentally alter the development lifecycle.

The first wave of research has largely focused on quantifying the productivity benefits of GenAI tools through short-term experiments and perception studies. These studies provide foundational evidence of productivity gains. For example, \citet{bird2023taking} found that developers using GitHub Copilot completed coding tasks substantially faster than their counterparts. Similarly, \citet{peng2023impact} observed that GenAI assistance enhances developer productivity. However, their primary focus on productivity remains limited. For them, productivity was primarily defined by task completion time and task success. This emphasis on completion time represents a partial view, focusing almost exclusively on developer output rather than on code quality or overall performance. 


Beyond simple speed metrics, these studies also touch upon other dimensions of productivity. \citet{peng2023impact} observed an increase in job satisfaction, suggesting that GenAI can reduce the burden of repetitive and frustrating tasks. However, this existing literature often consists of short-term experiments focused on individual, task-based scenarios \cite{bird2023taking, peng2023impact, imai2022github}. This narrow focus on singular metrics (e.g., task completion time) risks overlooking the broader team-level dynamics and the complex, multidimensional nature of developer productivity. Furthermore, significant challenges remain. The most immediate concerns involve the quality and reliability of AI-generated code, which can contain subtle bugs or vulnerabilities \citep{imai2022github, dakhel2023github}. This introduces a new cognitive burden: developers must shift effort from writing code to meticulously validating AI suggestions. This validation overhead, coupled with the tangible risk of over-reliance and skill atrophy \citep{bird2023taking, li2025hiding}, highlights the need for more holistic and longitudinal investigation.

The limitations of early studies underscore that developer productivity is not a singular metric but a complex, multidimensional construct. Measuring it effectively requires a holistic model that balances various factors. The SPACE framework \citep{forsgren2021space} provides such a model, defining productivity through five key dimensions: \textit{Satisfaction and well-being (S)}, \textit{Performance (P)}, \textit{Activity (A)}, \textit{Communication and collaboration (C)}, and \textit{Efficiency and flow (E)}. This framework is particularly well-suited to analyzing the sustained impact of GenAI. While the mentioned studies touched on \textit{Satisfaction} and \textit{Performance}, they did not capture the full picture. Our study addresses this gap by applying the SPACE framework to triangulate quantitative metrics with qualitative developer perceptions over time, providing a robust analysis of GenAI's long-term effects. The specific goal and research question guiding this longitudinal study are detailed in the following section.

\section{Case Study Design and Planning}
This section outlines the methodological foundation and planning of the study. We describe the study’s objectives, research question, cases, units of analysis, and subject selection, replication strategy, study design protocol, instrumentation, and data collection and analysis procedures. All the mentioned artifacts and analysis scripts are available in our online open science repository~\cite{zenodo}.




\subsection{Goal and Research Question}

Our investigation builds on the Goal–Question–Metric (GQM) paradigm \cite{basili2002tame}, which links analytical intent with measurable constructs: \textit{Analyze} the sustained adoption of GitHub Copilot and internal GPT tool \textit{for the purpose} of characterizing \textit{with respect to} their impact on software team productivity based on the SPACE framework \textit{from the point of view} of agile team members and researchers \textit{in the context of} a multi-case longitudinal study with agile teams from a multinational IT consulting firm. From this goal, we formulated the following Research Question (RQ): \textbf{\textit{How does the adoption of an internal GPT tool and GitHub Copilot impact the productivity of agile teams?}}. The rationale behind this question relies on the use of SPACE to examine how an internal GPT-based assistant and GitHub Copilot impact productivity in agile software teams. GenAI introduces cognitive and operational shifts that influence output, collaboration, and developers’ experience of flow and satisfaction. By decomposing productivity into different dimensions, the SPACE enables a broad understanding of how GenAI may transform both the human and technical dimensions of software work.

\subsection{Case Context, Units of Analysis, and Subject Selection}
\label{subsec:case_study_selection}
To answer our RQ, we adopted the industrial case study methodology, following the guidelines by 
\citet{runeson2012case}. This study was conducted in a large IT consulting firm, which we selected for two primary methodological reasons. First, it represents a critical case for productivity assessment. Its efficiency-driven business model, where teams operate under real deadlines, customer demands, and commercial pressure, makes it an appropriate context to observe productivity impacts. Second, it enables analytical generalization through heterogeneity \cite{yin2018case}. The firm's natural diversity across projects, customers, and technology stacks provides a representative setting for real-world software development. We gained access to three distinct projects, each with different customers, business domains, and team compositions. This industrial context provided an appropriate setting to evaluate the sustained impact of GenAI on professional practices, answering calls for more empirical validation in industry \cite{wohlin2012experimentation}.

The company’s agile delivery model (composed of \\semi-autonomous squads aligned with client projects) offered both methodological consistency and contextual diversity. The primary unit of analysis is the agile team (squad). The secondary unit is the individual developer, analyzed for perception and activity-level changes. This multi-level structure enables triangulation between individual experience and team productivity. From a practical standpoint, this evaluation was also designed to create a baseline for the firm, informing how GenAI impacts future project estimation and efficiency.





To ensure both methodological robustness and organizational feasibility in these real customer projects, a set of pre-selection criteria was established. These criteria balanced empirical requirements (e.g., data availability) and practical constraints (e.g., team engagement), enabling a reasoned selection of cases for longitudinal observation:

\begin{itemize}
\item \textit{Availability}: Projects with at least four completed sprints (historical data) and at least four additional sprints planned during the research period.

\item  \textit{Team Accessibility}: The extent to which project members were available and willing to engage in research activities, such as surveys, interviews, and workshops.

\item  \textit{Profile Capillarity}: Inclusion of diverse professional roles such as senior back-end developers focused on API integration, mid-level front-end engineers specializing in unit testing with Angular/React, and full-stack developers responsible for feature deliver, to enable the creation of representative personas.

\item \textit{Connectivity}: Verified access to the FM tools GitHub Copilot and internal GPT tool, ensuring participants could actively use and reflect upon these technologies.

\item \textit{Metrics/Traceability}: Technical feasibility of collecting relevant quantitative indicators from Jira, SonarQube, and Git repositories for cross-case comparison.
\end{itemize}

The studies were coordinated by a leadership team composed of professionals from both the customer-delivery teams and the firm’s innovation unit. Applying the aforementioned criteria, the leadership team identified ten potential projects, anonymized as Cases A–J, spanning diverse business domains: three from the banking sector (A–C), three from oil and gas (D–F), two from power and utilities (G–H), and two from logistics (I-J), one road-based and one maritime. Each of these projects satisfied the minimal requirement of four historical sprints and four future sprints planned within the study window. 

The case filtering process followed a pragmatic sequence. After conducting structured interviews with project managers and technical leads, five projects were retained for detailed evaluation. Cases E and H were excluded due to limited accessibility and low expected participant engagement, while C, G, and I were removed because telemetry extraction (e.g., from Jira and/or SonarQube) was infeasible in their environments.

During the longitudinal phase, two of the five remaining projects (A and B) were removed due to data access restrictions imposed by customer confidentiality protocols. The final study, therefore, comprised three completed cases (D, F, and J) representing two projects from the oil and gas domain and one from maritime logistics. Following the final case selection, we collaborated with project managers to identify the specific squads and professional profiles to be included in the analysis. These teams served as the units of analysis, and their members were characterized through the persona framework defined in the study protocol (see Section \ref{subsec:case_study}). This process ensured that each case reflected realistic professional diversity while maintaining conceptual comparability across contexts.

\subsection{Replication Strategy} 

The multi-case design followed theoretical replication logic as defined by Yin~\cite{yin2018case}. The intention was not to reproduce identical results across cases (literal replication), but to examine whether comparable patterns of impact (predicted by the SPACE framework) would manifest under different project and domain conditions (theoretical replication). 

In addition, replication was supported by a shared research protocol, which specified data collection procedures, metric definitions, and survey instruments. This protocol ensured that each team followed a consistent structure for sprint data extraction, survey timing, and feedback sessions. All participating professionals received a standardized onboarding and training program, including workshops on prompt engineering and GitHub Copilot usage, to minimize confounding effects caused by differing levels of familiarity with the tools.


\subsection{Study Design and Protocol}
\label{subsec:case_study}

\begin{figure}[ht]
  \centering
    \includegraphics[width=1\linewidth]{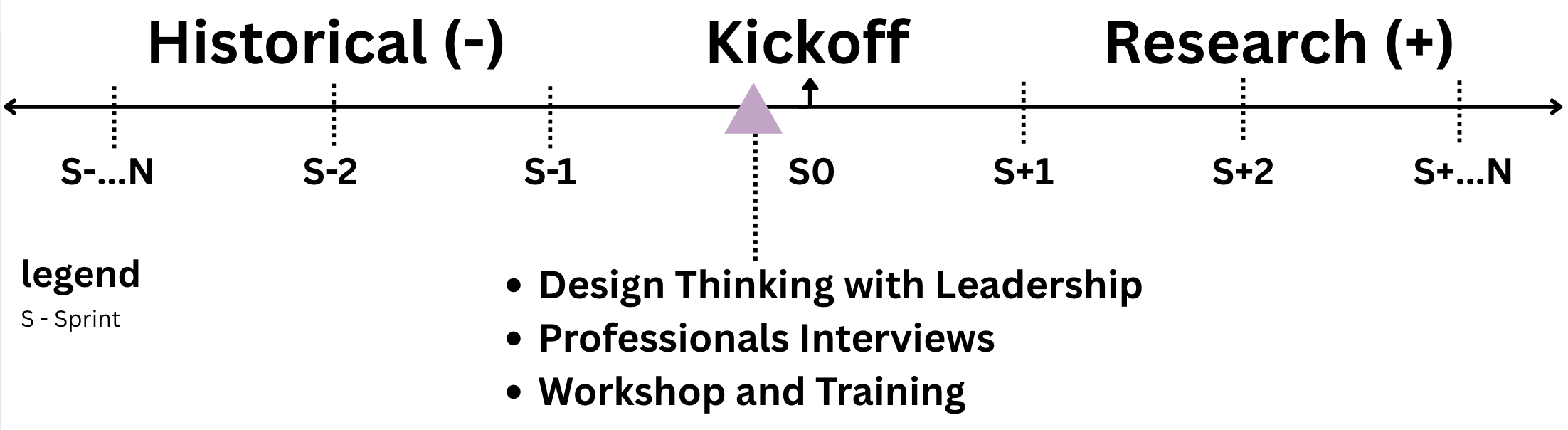}
  \caption{Case Study Protocol}
  \label{fig:timeline}
\end{figure}

As shown in Figure \ref{fig:timeline}, this multi-case longitudinal study was structured around two temporal phases: \textit{Historical Data} (before adoption) and \textit{Research Data} (after adoption), allowing a before-and-after comparison of behavioral and experiential change. Each case was observed for at least four and up to ten sprints per phase, long enough to capture two complete delivery cycles and to observe the stabilization of new practices following the introduction of the internal GPT tool and GitHub Copilot. The SPACE framework \cite{forsgren2021space} provided the conceptual lens for understanding multi-dimensional changes in productivity.



The protocol unfolded in five main stages, each designed to align with one or more SPACE dimensions:

\textbf{\textit{1. Historical Data Phase}}. The study began with the extraction of retrospective sprint data covering at least four completed iterations before GenAI adoption. These historical metrics established a baseline for productivity and quality through indicators such as story-point completion, sprint overflows, and lead time (Jira); code maintainability and vulnerability severity (SonarQube); and commit frequency and authorship traceability (GitHub/GitLab). Because the three projects began at different points in time, sprint sequences were normalized into relative time indices (S-N to S+N) to enable cross-case comparison. This phase captured the pre-adoption equilibrium of each team, the reference against which later effects would be assessed.


\textbf{\textit{2. Tool Introduction and Kickoff}}. The transition to the research phase began with a one-hour Kickoff workshop, during which all team members were introduced to the study goals, ethical procedures, and standardized practices for using the internal GPT tool and GitHub Copilot. A dedicated collaboration channel in Microsoft Teams was created for participants to exchange prompts, share experiences, and surface recurring challenges. This interactive space became a support mechanism.


\textbf{\textit{3. Research Data Phases}}. Once GenAI tools were fully integrated into daily work by the company's internal license and tools, the teams continued their regular agile routines across four to ten additional sprints. Throughout this period, data collection combined two complementary streams: 1) Automated quantitative metrics captured via Python scripts interfacing with Jira, SonarQube, and GitHub/GitLab APIs, covering throughput, quality, and activity indicators. 2) Perceptual surveys, administered twice per sprint, gathered developers’ evaluations of satisfaction, efficiency, cognitive load, and flow. Each Likert-scale item was paired with an open-ended question to contextualize numeric responses. When telemetry data were incomplete (such as missing Jira fields) metrics were normalized using available values and cross-checked against self-reported progress to maintain reliability. This triangulation allowed the study to document both what changed and how those changes were experienced.


\textbf{\textit{4. Design Thinking Sessions and Persona Definitions }}. Before the formal data collection phase, the consulting firm’s innovation unit and the research team conducted a series of Design Thinking sessions to co-design the study protocol and ensure its alignment with both organizational priorities and the research objectives. These sessions were led by the leadership team, which included researchers, project managers, and technical specialists representing the participating agile squads. The meetings followed a collaborative, iterative format typical of human-centered design approaches, combining structured facilitation with open discussion to balance methodological rigor and practitioner insight. Three key sessions guided this preparatory stage:

\begin{enumerate}
    \item \textit{Identification of Metrics and Indicators}: The first session focused on defining the key metrics necessary to operationalize the constructs of productivity, satisfaction, and well-being within the SPACE framework. Participants mapped metrics across three categories: technical (e.g., code quality, defect density), observability (e.g., velocity, sprint overflow, time per person), and business-oriented (e.g., value delivered per sprint), to ensure that data captured both engineering performance and perceived impact on day-to-day work.

\item \textit{Mapping of Technical Data Sources}: The second session concentrated on determining which engineering tools could reliably supply the data needed for quantitative analysis. The team validated the feasibility of extracting metrics from Jira, SonarQube, and GitHub/GitLab via API access, confirming that the required fields, update frequencies, and access permissions were available. This mapping exercise established the foundation for a unified and automated data collection pipeline across all cases.

\item \textit{Elicitation of Daily Challenges and Persona Definition}: The third session shifted focus from systems to people, exploring the everyday challenges and needs faced by software professionals in different roles. Through guided discussion and empathy mapping techniques, the team identified recurring “pain points” in areas such as testing, legacy maintenance, and feature integration. These insights informed the creation of personas, which are representative profiles such as a senior back-end developer focused on API integration, a mid-level front-end developer responsible for unit testing, and a full-stack engineer developing new features. The identified themes were prioritized according to relevance to the study goals: critical (7), intermediate (6), and non-critical (3). This prioritization ensured that subsequent surveys and interviews addressed issues most meaningful to the participants’ daily experience.

\end{enumerate}






\textbf{\textit{5. Professional Interviews and Training}}. Prior to the research phase, all participants took part in semi-structured interviews (conducted jointly with Microsoft specialists) to assess their familiarity with prompt engineering and GitHub Copilot. The interviews informed both the refinement of user personas and the design of a standardized training program to ensure comparable proficiency levels across teams. Every participant subsequently attended a two-hour remote workshop covering prompt-engineering concepts and practical demonstrations of Copilot usage. 



 

\subsection{Instrumentation}
\label{subsec:instrumentation}

Data collection instruments were designed to operationalize the five dimensions of the SPACE framework by combining automated telemetry, perceptual surveys, and qualitative validation procedures.


\textbf{\textit{Quantitative Instrumentation: API-Based Data Collection}}. A suite of Python scripts was developed to extract objective metrics directly from the engineering tools used in the participating projects. Each script accessed standardized API endpoints, providing traceable and reproducible measurements for the quantitative analysis:
\begin{itemize}
    \item \textit{Jira API — Performance (P)}: Extracted sprint-level metrics such as story points completed (classified as development or support tasks), sprint overflows, and lead time. These indicators capture throughput and delivery predictability.
    \item \textit{SonarQube API — Performance (P)}: Collected static analysis metrics including code vulnerabilities, duplication rates, and maintainability indices. These data reflect code quality as a proxy for sustainable performance over time.
    \item \textit{GitHub/GitLab API — Activity (A)}: Measured development activity through the number of commits, lines of code (LoC) added or deleted, and the traceability of commit authorship. These indicators capture both the rhythm and distribution of individual and team contributions.
\end{itemize}

All metrics were normalized by sprint length and team size to enable comparability across projects. To address incomplete telemetry data (e.g., missing Jira fields), we first imputed the missing values using available sprint averages. Subsequently, we verified data consistency by cross-checking these imputed values with developers’ self-reported progress.


\textbf{\textit{Perceptual Instrumentation: Developer Perception Survey}}. To complement the technical metrics, a developer perception survey was administered twice per sprint during the research phase. The instrument combined quantitative measures (5-point Likert scales) with qualitative feedback (open-ended comments) to capture the experiential aspects of GenAI adoption. Survey reliability was tested using Cronbach’s alpha ($\alpha$ = 0.96), confirming excellent internal consistency across the combined constructs of satisfaction, performance, and efficiency. 


\begin{itemize}
\item \textit{Satisfaction and Well-being (S)}: overall satisfaction with GenAI tools, perceived usefulness for specific tasks (e.g., API integration, legacy modernization), and sense of confidence in delivery.

\item \textit{Efficiency and Flow (E)}: perceived speed of delivery, reduction in repetitive or manual tasks, and diminished reliance on external information sources (e.g., online searches).

\item \textit{Performance (P)}: perceived assistance in increasing test coverage, creating new test scenarios, and improving feature development.
\end{itemize}



\textit{\textbf{Qualitative Instrumentation: - Interviews and Validation} - Communication and Collaboration (C)}. In the preparatory phase, semi-structured interviews (approximately 30 minutes each) were conducted with all participating professionals in collaboration with Microsoft specialists. These interviews served three purposes:
1) validating the relevance of the selected cases and confirming access to necessary data sources; 2) assessing each participant’s proficiency in prompt engineering and GitHub Copilot usage; and 3) informing the design of the standardized training workshop to ensure balanced skill levels across teams. Insights from these interviews also informed the persona definitions used in subsequent analysis, anchoring the human dimension of the study’s theoretical model.



\textit{\textbf{Training Workshop and Skill Standardization} - Communication and Collaboration (C)}. Following the interviews, all participants completed a two-hour online workshop on prompt engineering and effective GitHub Copilot use, conducted via Microsoft Teams. Training materials were drawn from Microsoft Reactor\footnote{https://developer.microsoft.com/en-us/reactor/} tutorials and from Microsoft Learn’s\footnote{https://learn.microsoft.com}. This preparation minimized learning-curve bias and reinforced methodological consistency across cases.

\textbf{\textit{Data Processing, Translation, and Anonymization}}. All raw data (quantitative and qualitative) were consolidated, pre-processed, and anonymized prior to analysis. A Python-based pseudonymization pipeline replaced identifiable attributes (e.g., usernames, project names, timestamps) with unique identifiers while preserving relational integrity across datasets (e.g., linking commits, survey responses, and sprints). Qualitative data were then translated and consolidated using Gemini v2.5 Flash with low-temperature inference parameters (temperature = 0.3, default top-p/top-k). This process supported linguistic consistency and facilitated thematic synthesis across multilingual data sources, while maintaining fidelity to the original meaning of participants’ responses.

\subsection{Data Collection Procedures}

Before data collection began, all participants were informed about the study’s goals, procedures, and ethical safeguards. Each developer consented to participate voluntarily, acknowledging that the investigation sought to evaluate the integration of GitHub Copilot and the internal GPT tool into their daily software development workflows. The study was also positioned internally as a reference for the consulting firm’s broader strategy to guide the responsible adoption of AI-assisted software engineering tools.

Data were collected over an 13-months period between October 2023 and November 2024, following a longitudinal multi-case design \cite{runeson2009guidelines}. Before the kickoff, employees were not permitted to use external GenAI tools like GitHub Copilot or Chat GPT for work. Following the kickoff, the company provided official licenses for the GitHub Copilot and released an internal GPT tool to all participants in the study group.
Each project was observed across two consecutive temporal phases: a Historical Phase (pre-adoption) establishing baseline metrics, and a Research Phase (post-adoption) capturing changes after the introduction of GenAI tools.
Each case contributed at least four historical sprints and four to seven research sprints, representing approximately eight to fourteen sprints per project.
This duration was intentionally defined to ensure trend stability and comparability while remaining aligned with each team’s natural release cadence, thereby capturing both the initial adaptation period and subsequent stabilization of GenAI-assisted practices.
Although projects started at slightly different times (see Table~\ref{tab:project_dates}), data were synchronized through sprint-based normalization to ensure equivalent observation windows across cases.




Project selection followed the procedure described in Section~\ref{subsec:case_study_selection}, resulting in three completed cases: Case D and Case F, both from the oil and gas sector, and Case J, from the maritime logistics sector. The final sample comprised 21 professionals representing diverse roles (front-end, back-end, and full-stack developers) and different levels of seniority (Jr, Mid, Sr). Each participant was linked to a representative persona defined during this phase, ensuring analytical coverage across different work contexts (Table~\ref{tab:personas_case_study_data}).


\begin{table*}[t]
    \centering
    \begin{tabular}{|c|c|c|c|c|c|c|c|c|}
        \hline
        \multirow{2}{*}{\textbf{Case}} & \multirow{2}{*}{\textbf{Squad}} & \multirow{2}{*}{\textbf{Sprint time box}} & \multicolumn{2}{c|}{\textbf{Historical}} & \multirow{2}{*}{\textbf{Kickoff}} & \multicolumn{2}{c|}{\textbf{Research}} & \multirow{2}{*}{\textbf{Number of Sprints}} \\
        \cline{4-5}
        \cline{7-8}
         &  &  & \textbf{Start} & \textbf{End} & & \textbf{Start} & \textbf{End} &  \\
        \hline
        \textbf{D} & \textbf{1} & 2 weeks & 30/10/2023 & 03/03/2024 & 04/03/2024 & 04/03/2024 & 23/06/2024 & 7 \\
        \textbf{F} & \textbf{1} & 2 weeks & 30/10/2023 & 03/03/2024 & 04/03/2024 & 04/03/2024 & 23/06/2024 & 7 \\
        \textbf{J} & \textbf{1} & 3 weeks & 13/11/2023 & 07/04/2024 & 01/04/2024 & 08/04/2024 & 01/09/2024 & 7 \\
        \textbf{J} & \textbf{2} & 3 weeks & 15/01/2024 & 09/06/2024 & 01/04/2024 & 10/06/2024 & 03/11/2024 & 7 \\
        \hline
    \end{tabular}
    \caption{Project Dates}
    \label{tab:project_dates}
\end{table*}

\begin{table}[!h]
    \centering
    \begin{tabular}{|c|c|c|c|c|}
    \hline
    \textbf{Profile} & \textbf{Role(dev)} & \textbf{Case} & \textbf{Squad} & \textbf{Persona} \\
    \hline
    profile\_01 & Backend/Mid & J & 2 & API+Kafka \\
    profile\_02 & Frontend/SR & J & 2 & Microfrontend \\
    profile\_03 & Fullstack/SR & F & 1 & Tests+Features \\
    profile\_04 & Backend/SR & J & 1 & API +Kafka \\
    profile\_05 & Frontend/JR & F & 1 & Tests\\
    profile\_06 & Frontend/SR & J & 2 & Microfrontend \\
    profile\_07 & Backend/SR & J & 2 & API+Kafka \\
    profile\_08 & Frontend/Mid & J & 1 & Microfrontend \\
    profile\_09 & Frontend/Mid & J & 2 & Microfrontend \\
    profile\_10 & Frontend/Mid & D & 1 & Legacy code+Bugs\\
    profile\_11 & Frontend/Mid & J & 1 & Microfrontend \\
    profile\_12 & Frontend/SR & J & 1 & Microfrontend \\
    profile\_13 & Backend/SR & J & 1 & API+Kafka \\
    profile\_14 & Backend/SR & D & 1 & Legacy code+Bugs \\
    profile\_15 & Fullstack/Mid & F & 1 & Tests+Features \\
    profile\_16 & Backend/Mid & J & 1 & API+Kafka \\
    profile\_17 & Frontend/JR & D & 1 & Legacy code+Bugs\\
    profile\_18 & Frontend/SR & J & 2 & Microfrontend \\
    profile\_19 & Frontend/SR & J & 2 & Microfrontend \\
    profile\_20 & Frontend/JR & J & 2 & Microfrontend \\
    profile\_21 & Backend/SR & J & 2 & API +Kafka \\
    \hline
    \end{tabular}
    \caption{Data Regarding Personas and Case Study}
    \label{tab:personas_case_study_data}
\end{table}



Data collection followed the protocol detailed in Section~\ref{subsec:case_study}, integrating automated quantitative telemetry with self-reported perceptual and qualitative data, all aligned to the SPACE framework \cite{forsgren2021space}.
This approach captured the multidimensional nature of productivity,
by linking what developers do (e.g., commits, story points, vulnerabilities) with how they experience their work (e.g., satisfaction, perceived flow, collaboration quality).

Triangulation was achieved through iterative cross-analysis between telemetry and survey data. Quantitative metrics were analyzed per sprint to identify productivity or quality trends, which were then compared against survey responses and open-text feedback.
Instances of divergence (e.g., increased story point completion but stagnant satisfaction levels) were flagged for deeper qualitative interpretation.
This methodological integration ensured construct validity by reconciling objective performance data with subjective developer experience, a key principle of the SPACE framework.
All instruments and scripts were prevalidated during this phase, ensuring the clarity of survey items, the reliability of automated extraction routines, and the feasibility of data access across corporate environments.

\subsection{Data Analysis Procedures}

The data analysis followed a structured and three-phase process combining descriptive, inferential, and qualitative techniques to examine both behavioral and perceptual changes associated with the use of the internal GPT tool and GitHub Copilot. All analytical procedures were explicitly guided by the SPACE framework and the study’s research question, ensuring theoretical and methodological coherence across data modalities.

\textbf{\textit{Phase 1: Data Preparation and Anonymization}}. The first analytical step focused on data integrity, privacy, and longitudinal traceability. All raw artifacts (including Git commit counter logs, Jira issue exports, and SonarQube reports) were consolidated by type into standardized DataFrames for each case. An anonymization algorithm (available in our open science repository \cite{zenodo}) was applied to remove identifiable attributes while preserving analytical relationships across datasets. Four structured mapping artifacts ensured a complete referential consistency:

\begin{itemize}
    \item  \textit{Users}: mapping original developer identifiers to anonymized profiles and case codes;
    \item \textit{Profile}: enriching anonymized profiles with role, seniority, and persona data;
    \item \textit{Case}: linking project identifiers to anonymized cases (D, F, J); and
    \item \textit{Period}: defining temporal segments for historical and research sprints.
\end{itemize}

Each artifact was validated for completeness and exported for quantitative and qualitative analyses.
This ensured privacy compliance and consistent cross-linking between metrics, roles, and time periods throughout the longitudinal analysis.

\textbf{\textit{Phase 2: Individual Artifact Analysis}}. Each anonymized artifact underwent independent analysis to detect within-case patterns and between-phase differences (Historical vs. Research). The goal was to identify changes in productivity, code quality, and developer activity following GenAI adoption. For each metric source — Jira (story points, sprint overflows), SonarQube (vulnerabilities, maintainability, code duplication), and GitHub/GitLab (commits, lines of code) — we computed descriptive statistics (mean, median, and standard deviation) and plotted temporal evolution using time-series visualizations.

To assess statistical significance between pre- and post-adoption phases, we applied the Mann–Whitney U test (for non-parametric paired samples).
Effect sizes (Cohen’s d) were computed to quantify the magnitude of observed differences, interpreted according to standard thresholds (d = 0.2 small, 0.5 medium, 0.8 large). Results emphasized trend convergence across cases rather than isolated p-values, ensuring balanced interpretation given the sample size.

Moreover, the perception survey dataset underwent psychometric validation before analysis. Internal consistency across constructs of Satisfaction and Well-being, Productivity, and Efficiency was verified using Cronbach’s alpha ($\alpha$ = 0.96), indicating excellent reliability. Likert-scale data were analyzed descriptively and inferentially using the same non-parametric and parametric tests described above, with results stratified by case, sprint, and persona.

On the other hand, open-ended survey responses were also analyzed. To synthesize this qualitative data, we employed Gemini v2.5 Flash.
The model processed comment fields grouped by case (D, F, J) and by SPACE-related constructs using the following standardized prompt:

\begin{figure}[h]
    \centering
    \definecolor{lightgray}{gray}{0.95}
    \definecolor{darkgray}{gray}{0.3}
    
    \fcolorbox{darkgray}{lightgray}{
        \begin{minipage}{0.95\columnwidth}
            \vspace{4pt}
            \small
            \textbf{Prompt Used for Qualitative Synthesis}
            \vspace{2pt}
            \hrule
            \vspace{6pt}
            
            \textit{Using the \texttt{form\_research.csv} file, filter the data for team cases (\texttt{team\_case\_anon}) D, F, and J. 
            Analyze the \texttt{Efficiency}, \texttt{Productivity}, and \texttt{Satisfaction} columns. 
            Group the information by \texttt{team\_case\_anon} and create bullet points for each metric (Efficiency, Productivity, Satisfaction) 
            based on the comment texts (\texttt{*\_comments}) and the \texttt{Context} and \texttt{Activity} columns.}
            \vspace{4pt}
        \end{minipage}
    }
    \caption{LLM Prompt for synthesis.}
    \label{fig:prompt}
\end{figure}


Asrecommended~\cite{baltes2025guidelines}, the LLM-generated synthesis was manually verified. Two researchers reviewed and reconciled the synthesis to ensure correctness. We use this qualitative analysis to support the reporting of our results. The complete reviewed summaries can be found in our online open science repository ~\cite{zenodo}.

\textbf{\textit{Phase 3: Cross-Case Synthesis and Triangulation}}. After completing within-case analyses, results were integrated through a cross-case synthesis following Yin’s replication logic \cite{yin2018case}.
Each case (D, F, J) was first analyzed independently to identify contextual trends, such as how domain complexity (oil and gas vs. logistics) influenced the adoption trajectory, and then compared across cases to examine pattern replication and divergence. Quantitative and qualitative data were triangulated iteratively.
Behavioral indicators (e.g., increased story points, reduced vulnerabilities) were cross-validated with perceptual trends (e.g., higher satisfaction or perceived efficiency).
Divergent patterns, such as productivity gains unaccompanied by improvements in well-being, were further explored through thematic analysis to identify underlying causes following the SPACE dimensions.

To ensure analytical rigor, different validity and reliability measures were embedded throughout the analysis. Researcher triangulation was applied as two analysts independently reviewed and reconciled qualitative syntheses to minimize interpretive bias. Temporal consistency was maintained through synchronized sprint-based comparisons, guaranteeing equivalent observation windows across all cases. Construct validity was reinforced by explicitly mapping each metric and interpretation to the SPACE framework, ensuring theoretical coherence between constructs and measures. Finally, statistical reliability was supported by combining significance testing with effect-size interpretation, prioritizing convergent trends across metrics and cases over isolated statistical anomalies.

\section{Study Results}

This section presents the findings of our longitudinal multi-case study examining how the adoption of GitHub Copilot and the intern GPT Tool impacted productivity among agile software teams in a global IT consulting firm. The analysis integrates quantitative telemetry from Jira, SonarQube, and Git repositories, utilizing the Likert scale, with qualitative and perceptual data from developer surveys and interviews conducted across three agile teams (Cases D, F, and J) over a 13-month period.



To address our research question, the analysis is structured according to the SPACE framework. This multidimensional approach enables a holistic characterization of productivity, moving beyond singular metrics to capture the interplay between developer experience, team performance, and development activity.   

\subsection{Satisfaction and Well-being (S)}

\begin{figure}[!h]
    \centering        
        \begin{subfigure}[b]{0.495\columnwidth}
            \centering
            \includegraphics[width=\linewidth]{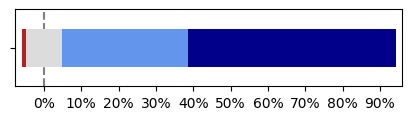}
            \caption{Satisfaction Assistant Role}
            \label{fig:form_research_satisfaction_assistant_role_likert}
        \end{subfigure}
        \hfill
        \begin{subfigure}[b]{0.495\columnwidth}
            \centering
            \includegraphics[width=\linewidth]{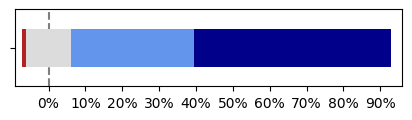}
            \caption{Overall Experience}
            \label{fig:form_research_satisfaction_overall_experience_likert}
        \end{subfigure}
        \hfill
        \begin{subfigure}[b]{0.495\columnwidth}
            \centering
            \includegraphics[width=\linewidth]{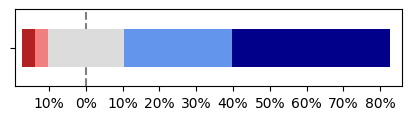}
            \caption{Unit Test Coverage - Overall}
              \label{fig:form_research_unit_tests_coverage_likert}
        \end{subfigure}
        \hfill
        \begin{subfigure}[b]{0.495\columnwidth}
            \centering
            \includegraphics[width=\linewidth]{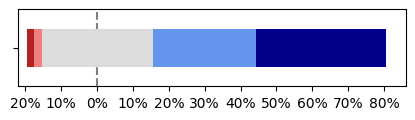}
            \caption{Unit Test Coverage - API}
            \label{fig:form_research_unit_tests_coverage_2_likert}
        \end{subfigure}
        \hfill
        \begin{subfigure}[b]{0.495\columnwidth}
            \centering
            \includegraphics[width=\linewidth]{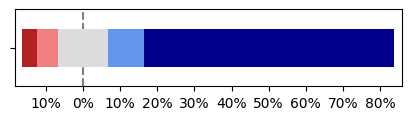}
            \caption{Unit Test Coverage - Front-End}
            \label{fig:form_research_unit_tests_coverage_3_likert}
            
        \end{subfigure}
        \hfill
        \begin{subfigure}[b]{0.495\columnwidth}
            \centering
            \includegraphics[width=\linewidth]{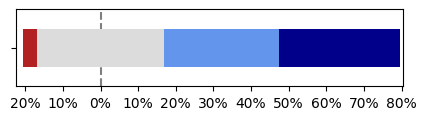}
            \caption{New Features Support}
            \label{fig:form_research_new_features_support_likert}
        \end{subfigure}
        \hfill
        \begin{subfigure}[b]{0.495\columnwidth}
            \centering  
            \includegraphics[width=\linewidth]{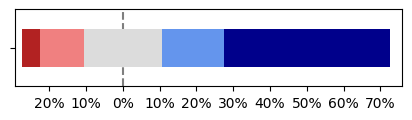}
            \caption{Test Scenarios Creating}
            \label{fig:form_research_test_scenarios_creation_likert}
        \end{subfigure}
        \hfill
        \begin{subfigure}[b]{0.495\columnwidth}
            \centering
            \includegraphics[width=\linewidth]{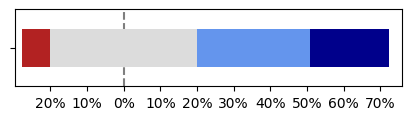}
            \caption{ APIs Support}
            \label{fig:form_research_apis_support_likert}
        \end{subfigure}
        \hfill
        \begin{subfigure}[b]{0.495\columnwidth}
            \centering
            \includegraphics[width=\linewidth]{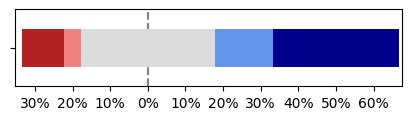}
            \caption{Legacy Code - JAVA}
            \label{fig:form_research_legacy_code_modernization_java_likert}
        \end{subfigure}
        \hfill
        \begin{subfigure}[b]{0.495\columnwidth}
            \centering
            \includegraphics[width=\linewidth]{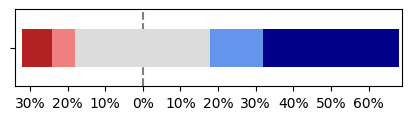}
            \caption{Legacy Code - REACT}
            \label{fig:form_research_legacy_code_modernization_jsf_react_likert}
        \end{subfigure}
        \hfill
        \begin{subfigure}[b]{0.495\columnwidth}
            \centering
            \includegraphics[width=\linewidth]{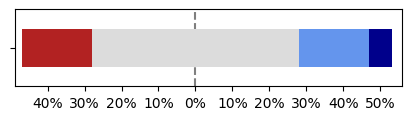}
            \caption{ETLs Migration Support}
            \label{fig:form_research_etls_migration_support_likert}
        \end{subfigure}
        \hfill
        \begin{subfigure}[b]{0.495\columnwidth}
            \centering
            \includegraphics[width=\linewidth]{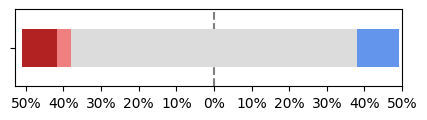}
            \caption{Kafka Support}
            \label{fig:form_research_kafka_support_likert}
        \end{subfigure}
    \hfill
        \begin{subfigure}[b]{1\columnwidth}
            \centering
            \includegraphics[width=\linewidth]{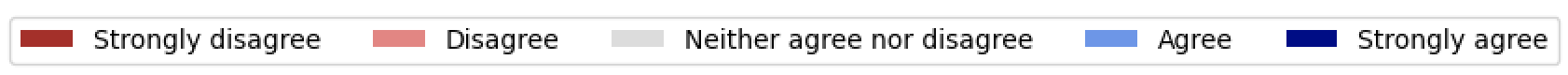}
            \label{fig:likert_scale}
            
        \end{subfigure}

        \caption{Perceived Satisfaction and Well-being (S)}
        \label{fig:form_research_likert}
 \end{figure}

The investigation into $S$atisfaction and Well-being reveals a strong and positive overall impact. Analysis of perceptual data shows a high median score for perceived well-being (3.78/5.0 mean, 4.0/5.0 median on a 5-point Likert scale), with $\approx$90\% of participants agreeing being satisfied with GenAI as an assistant and with their overall development experience (Figures \ref{fig:form_research_satisfaction_assistant_role_likert} and \ref{fig:form_research_satisfaction_overall_experience_likert}).   

However, this satisfaction was highly task-dependent. It was high for well-defined tasks like Unit Test Coverage ($\approx$75\% positive in Figures \ref{fig:form_research_unit_tests_coverage_likert}, \ref{fig:form_research_unit_tests_coverage_2_likert} and \ref{fig:form_research_unit_tests_coverage_3_likert}), and New Feature Development and Test Scenarios Creating ($\approx$65\% positive in Figures \ref{fig:form_research_new_features_support_likert} and \ref{fig:form_research_test_scenarios_creation_likert}). Conversely, satisfaction was exceptionally low for complex, context-heavy integration tasks, with a majority neutral or dissatisfied with support for APIs, both Legacy Code - JAVA and REACT ($\approx$50\% positive in Figures \ref{fig:form_research_apis_support_likert}, \ref{fig:form_research_legacy_code_modernization_java_likert} and \ref{fig:form_research_legacy_code_modernization_jsf_react_likert}) and ETL and Legacy Kafka Integration  (less than 25\% positive in Figures \ref{fig:form_research_etls_migration_support_likert} and \ref{fig:form_research_kafka_support_likert}). 

Qualitative feedback reinforced this differentiation. A senior developer (profile\_12) noted the tool is \textit{"really very useful"} for replicating test patterns \textit{"almost impeccably but it only works well for simple methods, for complex methods it fails to be reliable"}. This positive impact on well-being was also causally linked to a perceived increase in professional confidence. Participants also noted that Copilot \textit{"helps a lot in creating semantic commits... This helps in creating high-standard PRs"}. 

\subsection{Performance (P)}

The largest quantitative impact was observed in team $P$erformance. We used completed story points as a proxy for team output and SonarQube issues as a proxy for code quality.

Analysis of Jira data for Case J (which had consistent team membership) revealed a \textbf{59.1\% increase} in completed story points during the research period (from 281 to 447) compared to the historical baseline (Table \ref{tab:story_points_by_period}). This indicates a substantial rise in team throughput. Interestingly, the team's planned story points (from 447 to 1155) increased by $\approx$150\% in the same period, suggesting that GenAI adoption may have also increased the team's confidence in its capacity, leading them to commit to more work.

\begin{table}[h]
    \centering
    \begin{tabular}{|l|c|c|c|}
        \hline
        \textbf{Period} & \textbf{Planned} & \textbf{Completed} & \textbf{Remaining} \\
        \hline
        Historical & 447 & 281 & 166 \\
        Research & 1155 & 447 & 708 \\
        \hline
        \textbf{Total} & \textbf{1602} & \textbf{728} & \textbf{874} \\
        \hline
    \end{tabular}
    \caption{Story Points by Period and Status}
    \label{tab:story_points_by_period}
\end{table}

This change was statistically significant. We applied the \\Mann–Whitney U test and calculated the effect size using Cohen's d, comparing both the historical and research periods. For planned story points ($p-value$=4.94e-07, $d$=0.49) and remaining story points ($p-value$=6.15e-10, $d$=0.53), all returned $p-values$ well below the 0.05 threshold and medium effect size, suggesting a noticeable increase in story points during the research period compared to the historical period.  This result confirms that the introduction of GenAI tools had a statistically measurable impact on these key performance metrics.


The impact of this increased throughput on code quality presented a mixed picture. Analysis of SonarQube issues (Table \ref{tab:sonarqube_issues}) shows that during the research period, two of the three cases (F and J) substantially reduced their volume of 'High Severity' issues. Case J, for example, reduced its high-severity issues from 1624 to 825. Conversely, Case D experienced a degradation, with an increase in both 'High' and 'Low' severity issues. This finding suggests that, while some teams were able to improve quality while simultaneously increasing throughput, this positive outcome was not universal and may be context-dependent.

\begin{table}[h!]
    \centering
    \renewcommand{\arraystretch}{1.2} 
    \begin{tabular}{|l|l|c|c|c|c|}
    \hline
    \textbf{Period Type} & \textbf{Severity} & \multicolumn{3}{c|}{\textbf{Case}} & \textbf{\textit{Total}} \\
    \cline{3-5}
     & & \textbf{D} & \textbf{F} & \textbf{J} & \\
    \hline
    \multirow{2}{*}{Historical} & High Severity & 33 & 313 & 1624 & \textit{1970} \\
    \cline{2-6}
     & Low Severity & 28 & 29 & 1448 & \textit{1505} \\
    \hline
    \multicolumn{2}{|l|}{\textbf{Historical Sub Total}} & \textbf{61} & \textbf{342} & \textbf{3072} & \textbf{\textit{3475}} \\
    \hline
    \multirow{2}{*}{Research} & High Severity & 62 & 261 & 825 & \textit{1148} \\
    \cline{2-6}
     & Low Severity & 103 & 42 & 759 & \textit{904} \\
    \hline
    \multicolumn{2}{|l|}{\textbf{Research Sub Total}} & \textbf{165} & \textbf{303} & \textbf{1584} & \textbf{\textit{2052}} \\
    \hline
    \multicolumn{2}{|l|}{\textbf{\textit{Total}}} & \textbf{226} & \textbf{645} & \textbf{4656} & \textbf{\textit{5527}} \\
    \hline
    \end{tabular}
    \caption{SonarQube Issues of High and Low Severity by Case.}
    \label{tab:sonarqube_issues}
\end{table}

\subsection{Activity (A)}
The most critical finding emerged when gains in $P$erformance were contrasted with $A$ctivity. Analysis of 349 commit occurrences (totaling $\approx$456K lines) showed no statistically significant difference with p-value (p>0.05) ($p-value$=0.928), in the \textit{number\_of\_committed\_lines} variable between the historical and research periods, suggesting a negligible difference in the number of committed lines between the historical and research periods. We found a divergence between Performance (which rose sharply) and $A$ctivity (which remained flat). The teams achieved a 59.1\% increase in value ($P$, story points) with the same level of volume ($A$, number of committed lines). This decoupling between throughput (P) and volume of work (A) implies that productivity gains were not achieved through higher activity levels. Instead, developers delivered more value per line of code, suggesting a transformation toward value density rather than code quantity.

\subsection{Communication and Collaboration (C)}

The analysis of the Communication and Collaboration (C) was primarily qualitative, based on observations and feedback sessions. The process began with semi-structured interviews to assess initial proficiency and create personas. Following this, a standardized two-hour training workshop on effective Copilot use was conducted, minimizing learning bias and establishing a common vocabulary. These interventions had a positive impact by creating new formal (retrospectives) and informal (stand-ups, Microsoft Teams calls) forums for knowledge exchange and sharing. Furthermore, qualitative feedback suggested that GenAI shifted the focus of collaboration, with less time spent co-writing code and more time dedicated to review and integration. This transition suggests that GenAI did not reduce collaboration but rather redefined it (from shared construction to shared evaluation), thereby potentially preserving team cohesion while optimizing communication effort.

\subsection{Efficiency and Flow (E)}

\begin{figure}[!h]
    \centering
        \begin{subfigure}[b]{0.495\columnwidth}
            \centering
            \includegraphics[width=\linewidth]{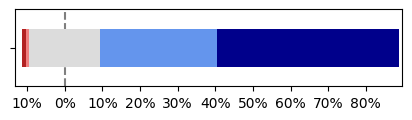}
            \caption{Avoiding Repetitive Work}
            \label{fig:form_research_efficiency_avoid_repetitive_tasks_likert}
        \end{subfigure}
        \hfill
        \begin{subfigure}[b]{0.495\columnwidth}
            \centering
            \includegraphics[width=\linewidth]{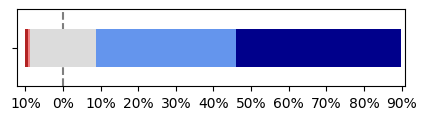}
            \caption{Delivery Speed}
            \label{fig:form_research_efficiency_delivery_speed_likert}
        \end{subfigure}

        \begin{subfigure}[b]{1\columnwidth}
            \centering
            \includegraphics[width=\linewidth]{figures/likert_scale.png}
            \label{fig:likert_scale}
            
        \end{subfigure}

        \caption{Efficiency }
        \label{fig:efficience_and_flow}
\end{figure}

Perceptual data from the survey data clearly explains that this efficiency increase stemmed from avoiding repetitive work and focusing on high-value tasks ($\approx$75\% in Figure \ref{fig:form_research_efficiency_avoid_repetitive_tasks_likert}) and in the perspective of the increased delivery speed ($\approx$81\% in Figure \ref{fig:form_research_efficiency_delivery_speed_likert}).

Qualitative comments directly linked this perceived efficiency to a reduction in cognitive friction and context switching. One developer (Profile 03) stated: "After using Copilot, the need for research in forums... was greatly reduced. GitHub Copilot really became my main source of reference." Another professional (Anonymous) confirmed this: "The fact of not needing to leave the IDE and its understanding of the context of my application helped a lot to solve problems... in little time."

\section{Discussion}

The longitudinal multi-case study provides in-situ evidence that extends beyond the scope of short-term experiments or isolated task evaluations. The integration of GenAI tools meaningfully reshapes productivity in agile software teams. The findings across the SPACE dimensions are interrelated, portraying a systemic transformation in developer work.



A primary finding is the divergence among $P$erformance, \\$E$fficiency, and $A$ctivity, a pattern we describe as the P–A–E divergence. Our results showed a 59.1\% increase in team $P$erformance, measured by completed story points. This finding provides real-world and longitudinal validation for the productivity gains reported in prior controlled studies (\textit{e.g.}, 55.8\% faster task completion~\cite{peng2023impact}). This quantitative leap was mirrored by our data on perceived $E$fficiency, where $\approx$82\% of participants reported increased speed and a reduction in repetitive tasks.

However, we found no statistically significant change in developer $A$ctivity ($p > 0.05$ for \textit{number\_of\_committed\_lines}). This finding, when contrasted with the surge in $P$erformance, suggests that GenAI's primary value is not in making developers ``do more" (\textit{i.e.}, increasing their raw activity), but in increasing the value density of their work. They achieved substantially greater throughput ($P$) with a similar volume of activity ($A$).

Qualitative evidence provides potential explanations for this $P-A-E$ divergence. Participants reported the main benefit was reducing ``search-and-synthesis" tasks and ``external research dependencies", effectively automating the ``toil" and cognitive friction that prior research has identified as a key drain on developer productivity. This aligns with the narrative of GenAI allowing developers to offload low-level implementation details and focus on higher-level problem-solving.

This shift directly contributed to our findings for the $S$atisfaction dimension. The high $S$-score (3.78/5.0 mean, 4.0/5.0 median) was explicitly linked to this reduction in toil and an "increased ability to solve challenges/bugs efficiently," which can support the developer in believing that their work has an impact \cite{storey2019towards}. 
This scaffolding of high-standard practices, which leds to ``greater confidence", suggests that GenAI may help mitigate the Impostor Phenomenon, a condition that \citet{guenes2024impostor} found to be an inhibitor of perceived productivity in software engineers. 



However, we found that gains were task-dependent. Some participants reported significant dissatisfaction with the tools on complex tasks like API and ETL integration and noted that the generated unit tests "required a lot of work to adapt". This aligns with findings from \citet{dakhel2023github} and \citet{peng2023impact}, confirming that the "validation overhead" remains a significant challenge. Participants also identified long-term risks, such as skill atrophy and "excessive tool dependence".

\begin{figure}[h]
    \centering
    \definecolor{lightgray}{gray}{0.95}
    \definecolor{darkgray}{gray}{0.3}
    
    \fcolorbox{darkgray}{lightgray}{
        \begin{minipage}{0.95\columnwidth}
            \vspace{4pt}
            \small
            \textbf{Key Takeaway}
            \vspace{2pt}
            \hrule
            \vspace{6pt}
            

            The adoption of GenAI tools in agile teams produced a measurable P–A–E divergence: team Performance (+59.1\%) and perceived Efficiency increased substantially, while developer Activity (commits, LoC) remained stable. These findings reveal that GenAI enhances productivity \textbf{not by intensifying labor through increased volume}, but by \textbf{concentrating its value density}. This shift enables developers to pivot from the friction of repetitive code generation toward high-leverage activities, such as architectural reasoning and rigorous validation.
            \vspace{4pt}
        \end{minipage}
    }
\end{figure}

\section{Threats to Validity}

While our longitudinal multi-case study design offers in-situ evidence, we acknowledge the following threats to validity, organized according to the categories recommended by~\citet{runeson2012case}.

Regarding \textbf{construct validity}, the potential divergence between our abstract constructs (e.g., productivity) and the metrics used to measure them, our perceptual measures for $S$atisfaction and $E$fficiency are susceptible to two main threats. First, participants' awareness of the study may have influenced their responses (the Hawthorne Effect). Second, the initial enthusiasm for using a novel technology (the Novelty Effect) may have temporarily inflated perceptions of satisfaction. It is unclear if these positive levels would be sustained after this novelty diminishes. 

Threats to \textbf{internal validity} concern the confidence with which we can claim a causal link between GenAI adoption and the observed outcomes. Our longitudinal design remains susceptible to maturation and history. A primary threat is that the 59.1\% productivity increase could be attributed to the teams' natural maturation and increased project familiarity over the 13-month study period. We note that this threat is partially mitigated by the fact that the participating professionals had already been working on their respective projects for at least one year before the study's kickoff, possessing substantial prior experience with the project context and codebase. Despite this, we cannot completely disentangle this effect from the intervention. This maturation threat is coupled with the tool's learning curve. An initial period of adjustment may have suppressed early productivity, just as later familiarity may have amplified it. These factors confound a direct causal attribution, as other unmeasured events, such as changes in project pressure, may have occurred concurrently.   

\textbf{External validity}, or the generalizability of our findings, is also a concern. Our findings originate from three teams within a single consulting firm, posing a threat of selection bias. Cases were selected based on pragmatic criteria, including data accessibility and willingness to cooperate. These teams may represent early adopters who are inherently more motivated, meaning the positive results may not generalize to teams with different cultural profiles or less openness to new technologies. Furthermore, although our longitudinal design is a strength, the 13-month period may not be sufficient for temporal generalizability. It may not capture long-term detrimental effects, such as technological dependence or a decrease in critical thinking, which participants flagged as a concern. The risk of skill atrophy remains a valid, long-term aspect that this study could not fully measure.

Finally, we employed several strategies to increase the \textbf{reliability} of our findings. First, the reliability of our perceptual survey was high, confirmed with a Cronbach's alpha of 0.96. Second, we standardized the intervention by providing a shared research protocol and a mandatory two-hour standardized training program to minimize learning-curve bias. Finally, as part of our protocol, we iteratively triangulated quantitative telemetry data (from Jira, Git, and SonarQube) with qualitative perceptual data (surveys and interviews) to ensure our interpretations were consistent and our constructs were valid.
\section{Conclusion}

This multi-case longitudinal study investigated the sustained impact of GenAI tools, specifically GitHub Copilot and an internal GPT tool, on three agile teams within a large IT consulting firm. By applying the SPACE framework, our approach provided an in-situ analysis that extends beyond short-term and task-based experiments.

The primary contribution of this work relates to its methodological context. While prior studies have mainly focused on short-term and individual experiments, our longitudinal, multi-case study complements them with empirical evidence of GenAI's sustained impact on established agile teams in a real-world, large IT consulting environment. Furthermore, the SPACE framework was adopted to capture a holistic view of productivity, moving beyond simplistic activity-based metrics. Doing so was important for uncovering the $P-A-E$ divergence, a nuanced finding invisible to studies measuring only task completion time or developer activity in isolation. Note that relying on $A$ctivity alone (number of committed lines) would have led to the false conclusion that GenAI had no impact.

For industry, this study offers additional evidence supporting the return on investment in generative AI adoption, demonstrating benefits that extend beyond simple speed-based metrics(\textit{e.g.}, we observed a positive overall impact on satisfaction and a 59.1\% increase in team-level story point throughput). For research, our findings open new directions for investigating the identified $P-A-E$ divergence and the observed task-dependent satisfaction patterns. 

Future work should include replications across diverse organizational and project contexts to evaluate the generalizability of these impacts and to analyze how different tools and task types (e.g., coding, testing, documentation) uniquely contribute to productivity outcomes.

\begin{acks}
We sincerely thank the Technology, Innovation, and Development team of the consultancy for their support and valuable contributions throughout this study. We are especially grateful to the project leaders for their guidance and encouragement. We also thank the Brazilian Research Council - CNPq (Grant 312275/2023-
4), Rio de Janeiro State's Research Agency - FAPERJ (Grant E-26/204.256/2024), the Coordination for the Improvement of Higher Education Personnel (CAPES), including the Postdoctoral Institutional Program (PIPD), the Kunumi Institute, and University of Bari TNE-DeSK project “Transnational Education Initiatives-Developing Shared Knowledge
in Smart Materials and Digital Transition for Sustainable Economy”, for their generous support.




\end{acks}

\bibliographystyle{ACM-Reference-Format}
\bibliography{sample-base}

@String{Computing = "Computing" }

@String{Springer = "Springer-Verlag" }

@article{ebert2023generative,
  title={Generative AI for software practitioners},
  author={Ebert, Christof and Louridas, Panos},
  journal={IEEE Software},
  volume={40},
  number={4},
  pages={30--38},
  year={2023},
  publisher={IEEE}
}

@article{sauvola2024future,
  title={Future of software development with generative AI},
  author={Sauvola, Jaakko and Tarkoma, Sasu and Klemettinen, Mika and Riekki, Jukka and Doermann, David},
  journal={Automated Software Engineering},
  volume={31},
  number={1},
  pages={26},
  year={2024},
  publisher={Springer}
}

@book{yin2018case,
  title={Case study research and applications},
  author={Yin, Robert K},
  volume={6},
  year={2018},
  publisher={Sage},
  address={Thousand Oaks, CA}
}

@article{bird2023taking,
  title={Taking flight with copilot},
  author={Bird, Christian and Ford, Denae and Zimmermann, Thomas and Forsgren, Nicole and Kalliamvakou, Eirini and Lowdermilk, Travis and Gazit, Idan},
  journal={Communications of the ACM},
  volume={66},
  number={6},
  pages={56--62},
  year={2023},
  publisher={ACM New York, NY, USA}
}

@article{dakhel2023github,
  title={Github copilot ai pair programmer: Asset or liability?},
  author={Dakhel, Arghavan Moradi and Majdinasab, Vahid and Nikanjam, Amin and Khomh, Foutse and Desmarais, Michel C and Jiang, Zhen Ming Jack},
  journal={Journal of Systems and Software},
  volume={203},
  pages={111734},
  year={2023},
  publisher={Elsevier}
}

@article{forsgren2021space,
  title={The SPACE of Developer Productivity: There's more to it than you think.},
  author={Forsgren, Nicole and Storey, Margaret-Anne and Maddila, Chandra and Zimmermann, Thomas and Houck, Brian and Butler, Jenna},
  journal={Queue},
  volume={19},
  number={1},
  pages={20--48},
  year={2021},
  publisher={ACM New York, NY, USA}
}

@article{peng2023impact,
  title={The impact of ai on developer productivity: Evidence from github copilot},
  author={Peng, Sida and Kalliamvakou, Eirini and Cihon, Peter and Demirer, Mert},
  journal={arXiv preprint arXiv:2302.06590},
  year={2023}
}

@inproceedings{imai2022github,
  title={Is github copilot a substitute for human pair-programming? an empirical study},
  author={Imai, Saki},
  booktitle={Proceedings of the ACM/IEEE 44th International Conference on Software Engineering: Companion Proceedings},
  pages={319--321},
  year={2022}
}

@book{wohlin2012experimentation,
  title={Experimentation in software engineering},
  author={Wohlin, Claes and Runeson, Per and H{\"o}st, Martin and Ohlsson, Magnus C and Regnell, Bj{\"o}rn and Wessl{\'e}n, Anders and others},
  volume={236},
  year={2012},
  publisher={Springer}
}

@article{runeson2009guidelines,
  title={Guidelines for conducting and reporting case study research in software engineering},
  author={Runeson, Per and H{\"o}st, Martin},
  journal={Empirical software engineering},
  volume={14},
  number={2},
  pages={131--164},
  year={2009},
  publisher={Springer}
}

@article{basili2002tame,
  title={The TAME project: Towards improvement-oriented software environments},
  author={Basili, Victor R and Rombach, H Dieter},
  journal={IEEE Transactions on software engineering},
  volume={14},
  number={6},
  pages={758--773},
  year={2002},
  publisher={IEEE}
}

@inproceedings{guenes2024impostor,
  title={Impostor phenomenon in software engineers},
  author={Guenes, Paloma and Tomaz, Rafael and Kalinowski, Marcos and Baldassarre, Maria Teresa and Storey, Margaret-Anne},
  booktitle={Proceedings of the 46th International Conference on Software Engineering: Software Engineering in Society},
  pages={96--106},
  year={2024}
}

@inproceedings{gandhira2025adoption,
  title={Adoption of Agile Scaling Practices in Globally Distributed Software Development Teams: A Systematic Literature Review},
  author={Gandhira, Sajani and Wickramarachchi, Ruwan},
  booktitle={2025 International Research Conference on Smart Computing and Systems Engineering (SCSE)},
  pages={1--6},
  year={2025},
  organization={IEEE}
}

@article{assalaarachchi2025generative,
  title={Generative AI for Software Project Management: Insights from a Review of Software Practitioner Literature},
  author={Assalaarachchi, Lakshana Iruni and Masood, Zainab and Hoda, Rashina and Grundy, John},
  journal={IEEE Software},
  year={2025},
  publisher={IEEE}
}

@article{murali2024ai,
  title={AI-assisted Code Authoring at Scale: Fine-tuning, deploying, and mixed methods evaluation},
  author={Murali, Vijayaraghavan and Maddila, Chandra and Ahmad, Imad and Bolin, Michael and Cheng, Daniel and Ghorbani, Negar and Fernandez, Renuka and Nagappan, Nachiappan and Rigby, Peter C},
  journal={Proceedings of the ACM on Software Engineering},
  volume={1},
  number={FSE},
  pages={1066--1085},
  year={2024},
  publisher={ACM New York, NY, USA}
}

@article{zhang2023practices,
  title={Practices and challenges of using github copilot: An empirical study},
  author={Zhang, Beiqi and Liang, Peng and Zhou, Xiyu and Ahmad, Aakash and Waseem, Muhammad},
  journal={arXiv preprint arXiv:2303.08733},
  year={2023}
}

@article{wiggins2022opportunities,
  title={On the opportunities and risks of foundation models for natural language processing in radiology},
  author={Wiggins, Walter F and Tejani, Ali S},
  journal={Radiology: Artificial Intelligence},
  volume={4},
  number={4},
  pages={e220119},
  year={2022},
  publisher={Radiological Society of North America}
}

@misc{zenodo,
  author={{Anonymous}},
  title={{Impacts of Generative AI on Agile Teams' Productivity: A Multi-Case Longitudinal Study [Data set]}},
  howpublished={{Zenodo}},
  year={{2025}},
  note={{Associated with: 3rd ACM International Conference on AI Foundation Models and Software Engineering (FORGE 2026), Rio de Janeiro, Brazil}},
  doi={10.5281/zenodo.17536708},
  url={{https://doi.org/10.5281/zenodo.17536708}}
}

@article{li2025hiding,
  title={When Hiding Hurts: How Stealth Use of Generative AI Resources Impairs Task Performance},
  author={Li, Peikai and Ma, Lin and Dong, Niannian and Xing, Lu Lucy and Yuan, Shuai},
  journal={Shuai, When Hiding Hurts: How Stealth Use of Generative AI Resources Impairs Task Performance (January 03, 2025)},
  year={2025}
}

@book{runeson2012case,
  title={Case study research in software engineering: Guidelines and examples},
  author={Runeson, Per and Host, Martin and Rainer, Austen and Regnell, Bjorn},
  year={2012},
  publisher={John Wiley \& Sons}
}

@article{baltes2025guidelines,
  title={Guidelines for Empirical Studies in Software Engineering involving Large Language Models},
  author={Baltes, Sebastian and Angermeir, Florian and Arora, Chetan and Bar{\'o}n, Marvin Mu{\~n}oz and Chen, Chunyang and B{\"o}hme, Lukas and Calefato, Fabio and Ernst, Neil and Falessi, Davide and Fitzgerald, Brian and others},
  journal={arXiv preprint arXiv:2508.15503},
  year={2025}
}

@article{storey2019towards,
  title={Towards a theory of software developer job satisfaction and perceived productivity},
  author={Storey, Margaret-Anne and Zimmermann, Thomas and Bird, Christian and Czerwonka, Jacek and Murphy, Brendan and Kalliamvakou, Eirini},
  journal={IEEE Transactions on Software Engineering},
  volume={47},
  number={10},
  pages={2125--2142},
  year={2019},
  publisher={IEEE}
}










\end{document}